\documentclass[12pt]{spieman}  
\usepackage{amsmath,amsfonts,amssymb}
\usepackage{graphicx}
\usepackage{setspace}
\usepackage{tocloft}
\usepackage{lineno}
\linenumbers
\title{Single-shot, simultaneous spatially resolved polarimetry and wavefront sensing with stress engineered optics}

\author[a*]{David Spiecker}
\author[a]{Thomas Brown}

\affil[a]{Institute of Optics, University of Rochester, 480 Intercampus Dr, Rochester, NY 14627, USA\\}

\cftpagenumbersoff{figure}
\cftpagenumbersoff{table} 
\begin{document} 
\nolinenumbers
\maketitle

\begin{abstract}
We present an experimental test of the use of stress engineered optics incorporated into a Shack-Hartmann wavefront sensor in such a way that the shape of the point spread function (PSF) provides polarization information while the displacement of the PSF gives information about the local wavefront gradient, allowing a reconstruction of the wavefront gradient. Using calibrated polarizations and wavefront tilts we construct a measurement matrix capable of reconstructing both the Stokes parameters and the wavefront, yielding a measurement of wavefront gradients as small as 100 $\mu$rad with a polarimetric angular error of approximately 0.1 radians on the Poincar\'e sphere in a single frame measurement.     
\end{abstract}

\keywords{polarimetry, wavefront sensing, stress engineered optics, stress birefringence, Shack-Hartmann wavefront sensor}

{\noindent \footnotesize\textbf{*}David Spiecker,  \linkable{davidspiecker@rochester.edu} }

\begin{spacing}{2}   

\section{Introduction}
\label{sect:intro}
Accurate and reliable single-shot or single frame measurements of optical fields are of great importance both in high power laser technology and in systems under motion in which drift and vibration could induce errors. Optical fields must be characterized primarily through intensity or irradiance measurements; measurements of relative phase and polarization must therefore be measured indirectly such that the irradiance measurement carries phase or polarization information within it.  Single frame measurements of wavefront/phase exist in the form of Shack-Hartmann sensors (in high-energy lasers, astronomy, ophthalmology, and adaptive optics\cite{schwiegerling_historical_2005}) and division of wavefront interferometers while similar methods have been applied to the measurement of polarization. In particular, Shack-Hartmann sensors have been vital in adaptive optics\cite{williams_wavefront_1999} and in the characterization of high powered lasers.\cite{kruschwitz_accurate_2012} One of the earliest single-shot polarimeters was described in 1985 by Azzam\cite{azzam_arrangement_1985} using a four-detector arrangement.  Since then, it has been recognized that the large number of pixels available on an image sensor can, in a suitably calibrated system, yield polarization-sensitive images and accurate Stokes polarimetry. \cite{kemme_snapshot_2007,oka_compact_2003,li_broadband_2018,chun_polarization-sensitive_1994,zimmerman_star_2016}

In this paper, we describe progress toward the design of a single-shot Stokes polarimeter combined with a Shack-Hartmann sensor.  The concept is based on the realization that the spot array created by the lenslets in a Shack-Hartmann sensor naturally fit with the idea of a star test, in which the properties of the field may be deduced from the shape of a point spread function (PSF).  The two key optical elements in the system are a stress-engineered optic (SEO) and a lenslet array, both which enable polarimetry and wavefront sensing, respectively. SEOs have been applied to image-sampling Stokes polarimetry in prior work by Zimmerman and Brown \cite{zimmerman_star_2016} and to sampling the Stokes parameters in a multicore fiber by Sivankutty et al.\cite{sivankutty_single-shot_2016} In these systems, an SEO is most easily implemented as plane-parallel window with an external, peripheral stress distribution that in the simplest cases, has trigonal symmetry.
\begin{figure}
    \centering
    \includegraphics[width=0.85\linewidth]{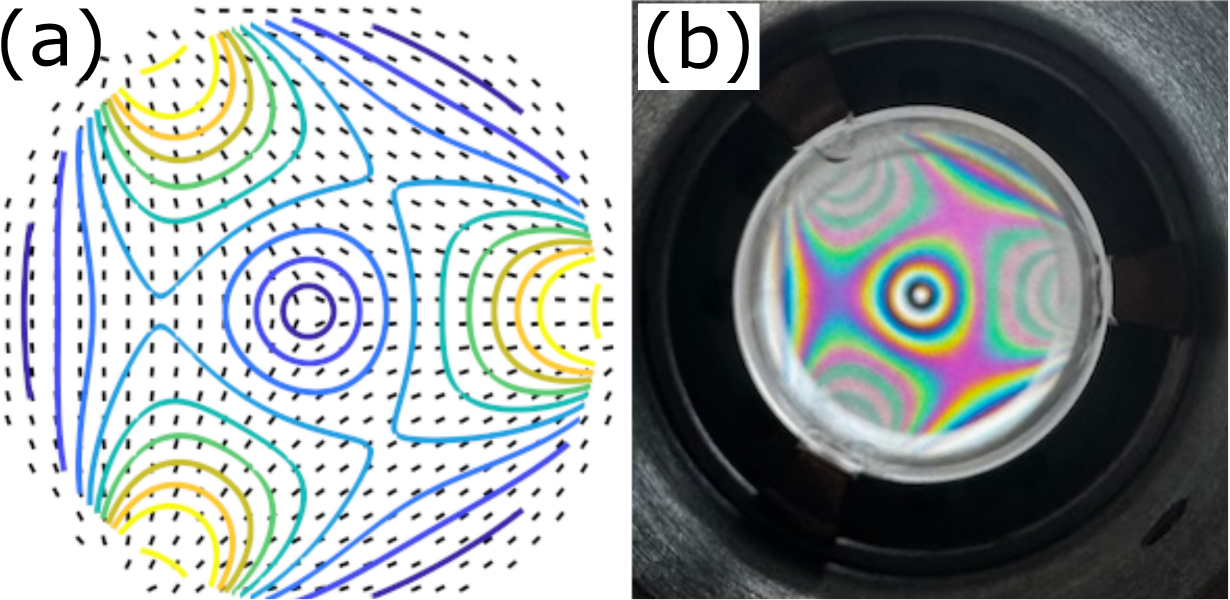}
    \caption{(a) Theoretical model of an SEO, where the short, black lines indicates the slow axis orientation and the colored lines indicate contours of equal retardance in an SEO. (b) A physical SEO viewed with a right-circular polarized input and a matching analyzer, demonstrating the phase retardance patterns present in SEOs. The phase retardance pattern presents itself through the observed intensity of light and the phase retardance in the central region varies linearly.}
    \label{fig:theory_exp_model_SEO}
\end{figure}

\subsection{Polarimetry with stress engineered optics}
The behavior of stress in an SEO is well described using the theoretical model developed by Yiannopoulos \cite{yiannopoulos_general_1999} and applied to SEOs by Brown and Beckley.\cite{brown_stress_2013} With the knowledge of the stress distribution in an optical window, the slow axis orientation and the phase retardance at every point in the window can be predicted. Figure \ref{fig:theory_exp_model_SEO} shows a typical result of such a calculation assuming threefold symmetry with external stress distributed tightly over three regions separated by 180$^\circ$. A key parameter of SEOs is the dimensionless stress parameter $c$ which describes the rate of change of the phase retardance over normalized radial distances in the central region of SEOs. The dimensionless stress parameter is defined through the phase retardance in an optical element, which is defined as 
\begin{equation}
    \delta(\rho,\phi)= \frac{ 2\pi}{\lambda}t\Gamma(\rho,\phi),
    \label{eq:phase_retardance}
\end{equation}
where $\rho$ is the normalized (to the radius of the optical window) radial direction and $\phi$ is the azimuthal direction in a cylindrical polar coordinate system, $t$ is the thickness of the SEO in the direction of the propagation of light, $\lambda$ is the wavelength of the light, and $\Gamma$ is the stress birefringence of the medium. The stress birefringence depends on the loading geometry of the optical window and the window's material properties. Of particular interest is the region near the center of the SEO (up to approximately $\rho=0.2$). In this region, the phase retardance varies only in the radial direction and the fast axis orientation varies only in the azimuthal direction. For SEOs with trigonal loading geometry such as the one used in our experiments, the phase retardance varies linearly and can be expressed as 
\begin{equation}
    \delta(\rho)=c\rho,
    \label{eq:retardance_power_law}
\end{equation}
where $c$ is the dimensionless stress parameter which is proportional to the external applied force and represents the rate of change of the phase retardance over normalized radial distances.\cite{brown_stress_2013}

\begin{figure}[h]
	\includegraphics[width=0.8\linewidth]{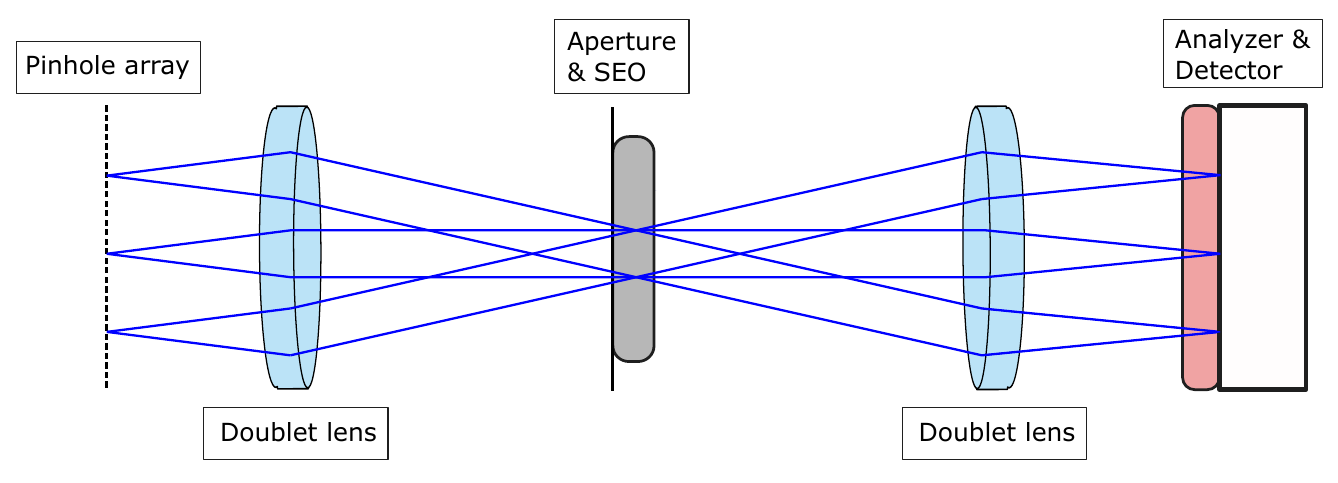} 
	\centering
	\caption{Star Test Image Sampling Polarimeter as developed by Zimmerman and Brown in which the pinhole array creates an array of point sources sampling a scene of interest.\cite{zimmerman_star_2016}}
	\label{fig:STIP}
\end{figure}

\begin{figure}[h]
    \centering
    \includegraphics[width=0.8\linewidth]{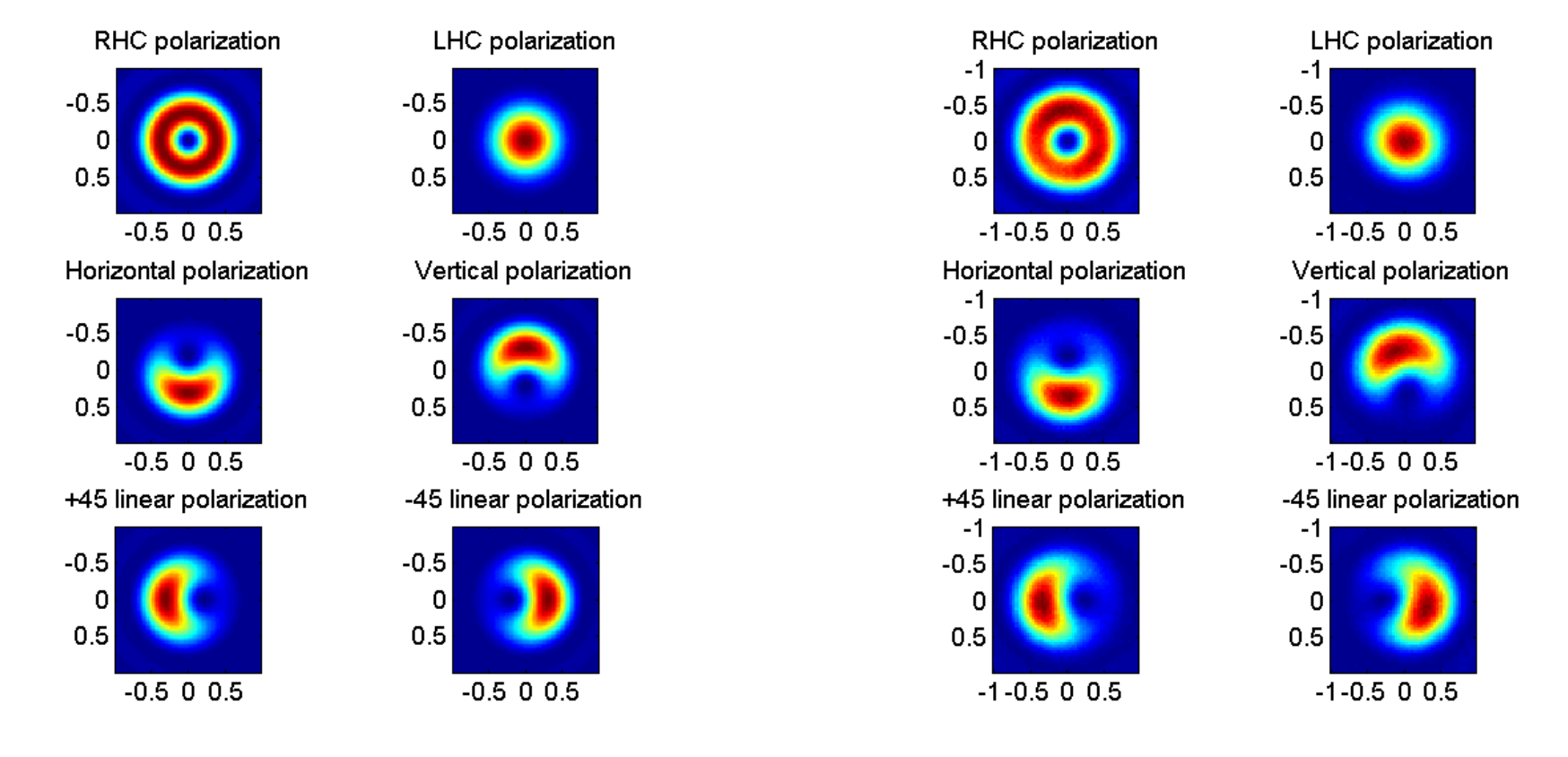}
    \caption{PSFs of reference polarization states as predicted (left) and measured (right) by Ramkhalawon. (Reproduced with permission from Ramkhalawon, Brown, and Alonso\cite{ramkhalawon_star_2012})}
    \label{fig:SEO_PSFs}
\end{figure}

When the SEO is placed in the Fourier plane of a 4F imaging system (Fig. \ref{fig:STIP}) and a point source is imaged, the resulting PSF (with a polarization analyzer) acquires an irradiance pattern that is unique to the input polarization state as demonstrated in Fig. \ref{fig:SEO_PSFs}. \cite{ramkhalawon_star_2012,ramkhalawon_imaging_2013}
The polarization state of an input field are represented by the Stokes parameters $S_0, S_1, S_2,$ and $S_3$. The polarization states described by the Stokes parameters $S_1, S_2,$ and $S_3$ are referenced as H, V, P, M, R, L for horizontal, vertical, +45, -45, right-circular, and left-circular respectively. Since the irradiance pattern at the detector is uniquely related to the input polarization, the system can be modeled with a measurement matrix $\mathbf{M}$ as shown in Eq. (\ref{eq:measurement_matrix_equation}),
\begin{equation}
    I(\vec{x})=\mathbf{M}(\vec{x})\vec{S},
    \label{eq:measurement_matrix_equation}
\end{equation}
where $I$ is the irradiance pattern, $\vec{x}$ is a vector of all image points, and $\vec{S}$ is the Stokes vector representing the input polarization state. The measurement matrix (a $j\times 4$ matrix where $j$ is the number of elements in $\vec{x}$) can be constructed from the product of the pseudoinverse of the matrix of N concatenated Stokes vectors (represented as a $4 \times N$ matrix with each column representing an input polarization state) and their corresponding irradiance patterns (represented as a $j \times N$ matrix) as shown in Eq. (\ref{eq:measurement_matrix_construction}).

\begin{equation}
    \mathbf{M}(\vec{x})=[I_{\vec{x},1}, \; I_{\vec{x},2}, \; \cdots , \; I_{\vec{x},N}] 
    \begin{bmatrix}
        S_{0,1}, \; S_{0,2}, \; \cdots , \; S_{0,N} \\
        S_{1,1}, \; S_{1,2}, \; \cdots , \; S_{1,N} \\
        S_{2,1}, \; S_{2,2}, \; \cdots , \; S_{2,N} \\
        S_{3,1}, \; S_{3,2}, \; \cdots , \; S_{3,N} \\
    \end{bmatrix}^{-1}
    \label{eq:measurement_matrix_construction}
\end{equation}
This measurement matrix is constructed with a monochromatic input. The unknown Stokes vector $\vec{S}_{U}$ may be retrieved from the measurement of an irradiance pattern if $\mathbf{M}(\vec{x})$ is known. By performing a pseudoinverse operation on the measurement matrix to obtain a data reduction matrix $\mathbf{M}(\vec{x})^{-1}$, the polarization state of the input can be retrieved from the product of the data reduction matrix and the measured irradiance pattern $I_{meas}(\vec{x})$ as expressed by Eq. (\ref{eq:Stokes_retrieval}).
\begin{equation}
   \vec{S}_{U}=\mathbf{M}(\vec{x})^{-1} I_{meas}(\vec{x})
    \label{eq:Stokes_retrieval}
\end{equation}

\begin{figure}
    \centering
    \includegraphics[width=0.5\linewidth]{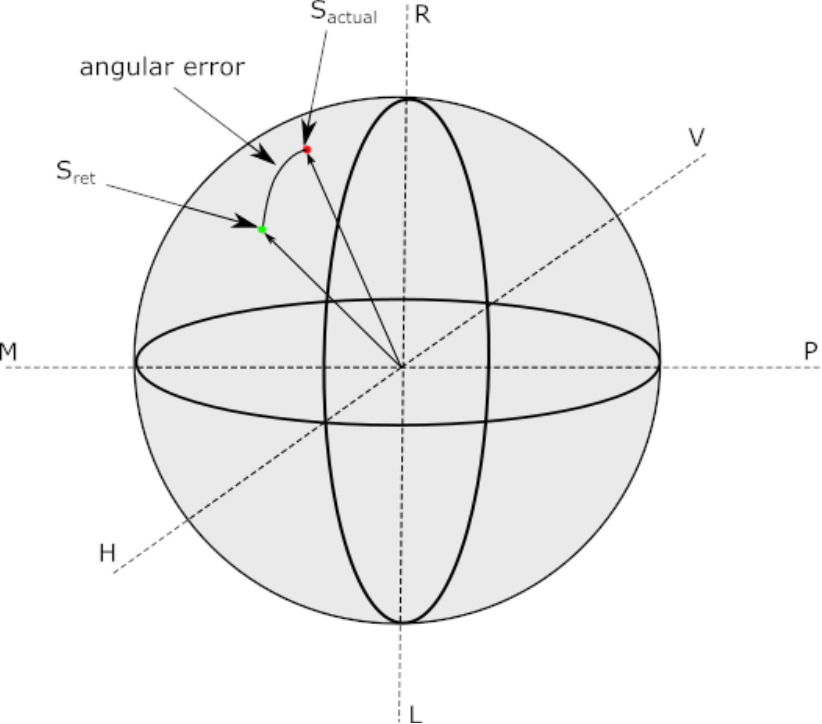}
    \caption{Angular error as illustrated on the Poincaré sphere}
    \label{fig:angular_error}
\end{figure}

The performance of the system is determined by the polarimetric angular error. The polarimetric angular error is defined as the angular separation of the input's Stokes vector and the retrieved Stokes vector as represented on the Poincaré sphere (Fig. \ref{fig:angular_error}).

\subsection{Wavefront sensing with a Shack-Hartmann wavefront sensor}
A Shack-Hartmann wavefront sensor (SHWFS) is a wavefront sensing technique that spatially resolves an input wavefront's gradient. The SHWFS utilizes a lenslet array where each lenslet focuses a small region of an input wavefront onto a detection plane (at a distance of the lenslet's focal length, the detection plane overlaps the lenslet's image plane). In the case of an ideal, non-aberrated wavefront, each lenslet focuses the input wavefront to a spot at the intersection of its optical axis and the image plane. For a slowly varying aberrated wavefront, the local gradient is approximately linear for a sufficiently small lenslet. A linear gradient in the wavefront corresponds to a displaced focal spot from the optical axis on the image plane when focused by the lenslet. 

\begin{figure}
    \centering
    \includegraphics[width=0.7\linewidth]{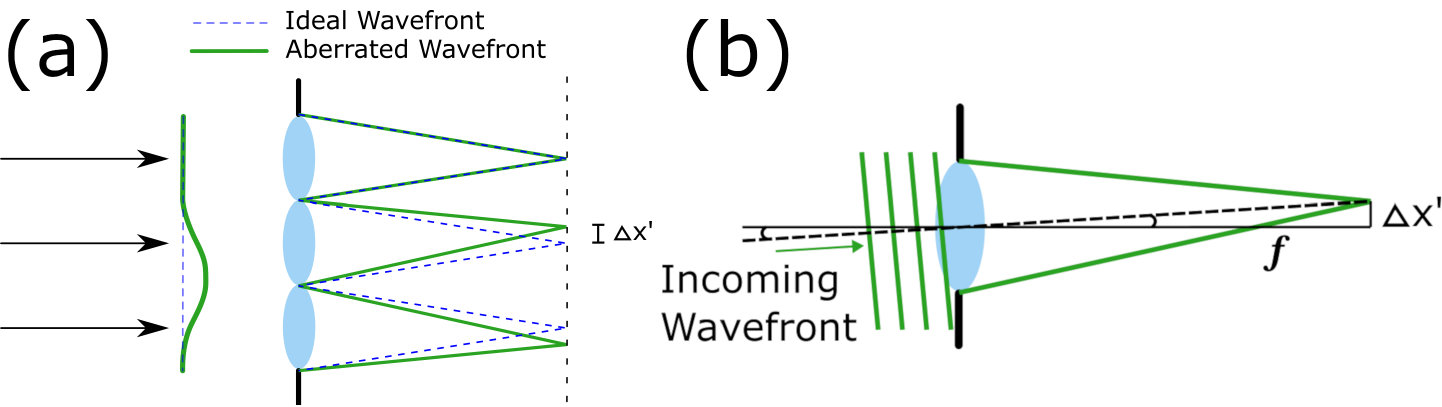}
    \caption{An illustration of the relationship between a wavefront that is incident on a lenslet array and the resulting spot position. (a) Comparison between an ideal wavefront and an aberrated wavefront and changes to spot locations. (b) The relationship between the wavefront gradient incident onto a lenslet and the resulting spot displacement.}
    \label{fig:SHWFS_wavefront}
\end{figure}

The relationship between the wavefront gradient and the displacement of the focal spot is expressed as 
\begin{equation}
    \frac{\partial{W}(x,y)}{\partial{x}}=-\frac{\Delta x'}{f},
    \label{eq:wavefront gradient}
\end{equation} where $\partial{W}/\partial{x}$ is the wavefront gradient along the x-direction, $\Delta x'$ is the spot displacement in the x-direction at the image plane, and $f$ is the lenslet's focal length. By measuring the location of the focal spot and its displacement from the optical axis, the gradient of the wavefront incident at each lenslet can be retrieved as illustrated in Fig. \ref{fig:SHWFS_wavefront}.
At its simplest, the retrieval algorithm is a center-of-mass centroiding\cite{neal_shack-hartmann_2002}. There are also more sophisticated algorithms that allow for more precise position retrieval by correlating the detected image with an expected image.\cite{poyneer_scene-based_2003,poyneer_experimental_2005}

Once the wavefront gradient is extracted from Shack-Hartmann data, the wavefront may be reconstructed using a variety of techniques. Most of those techniques use a set of orthogonal polynomials to describe the reconstructed wavefront.\cite{noll_zernike_1976, southwell_wave-front_1980,lane_wave-front_1992} In adaptive optics and Shack-Hartmann sensing, Zernike polynomials are used to represent the wavefront $W$, 
\begin{equation}
    W(\rho,\theta) = \sum_n a_nZ_n(\rho,\theta),
    \label{eq:Wavefront_Zernike}
\end{equation}
where $\rho$ and $\theta$ are normalized polar coordinates, $a_n$ are the Zernike coefficients, and $Z_n$ are the $n$th Zernike polynomial (in Noll indexing). The wavefront gradient is determined by taking the spatial derivative of Eq. \ref{eq:Wavefront_Zernike}, or by using the chain rule, the derivative with respect to normalized spatial coordinates as shown by,
\begin{equation}
    \nabla W(\rho,\theta) = \sum_n a_n \nabla Z_n(\rho,\theta) = \sum_n a_n \frac{1}{R}\nabla_{\rho} Z_n(\rho,\theta),
    \label{eq:Wavefront_Zernike_derivative}
\end{equation}
where $R$ is the radius of the aperture used to define the area the polynomial defines. Equation \ref{eq:Wavefront_Zernike_derivative} can be rearranged into 
\begin{equation}
    \frac{R}{\lambda}\nabla W(\rho,\theta) = \sum_n a_n \nabla_{\rho} Z_n(\rho,\theta),
    \label{eq:Zernike_coeff_equation}
\end{equation}
where it can be reinterpreted using linear algebra such that $a_n$ is a coefficient matrix and the remaining terms are vectors. The coefficient matrix can be determined as the least-squares solution to the equation if the wavefront gradient is known. 

\subsection{Single-shot, simultaneous spatially resolved polarimetry and wavefront sensing}
\begin{figure}
	\centering
	\includegraphics[width=0.9\linewidth]{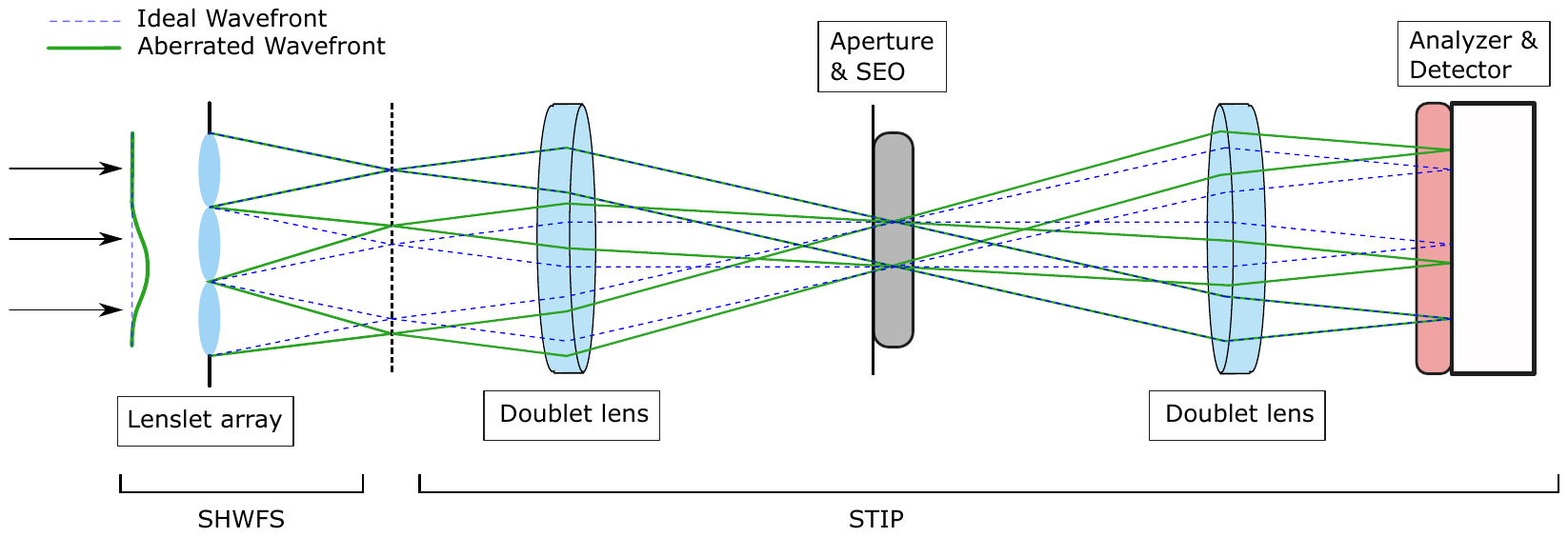} 
	\caption{SHWFS-STIP schematic}
	\label{fig:SHWFSSTIP}
\end{figure}

A SHWFS creates an array of point sources at its image plane and a STIP requires an array of point sources as an input. By overlapping the SHWFS image plane and the STIP object plane, it becomes possible to simultaneously perform polarimetry and wavefront sensing with a single image. A schematic of this set up is shown in Fig. \ref{fig:SHWFSSTIP}.
The spot displacement produced by the SHWFS with an aberrated wavefront as an input is preserved by the STIP. As the input wavefront propagates through the STIP, each spot acquires a polarization-dependent irradiance pattern. At the detection plane of the SHWFS-STIP, each spot presents a polarization-dependent irradiance pattern and displacement corresponding to the wavefront gradient. The SHWFS-STIP can be modeled with a similar method as Eq. \ref{eq:measurement_matrix_equation} where an irradiance pattern is related to the input's Stokes vector and wavefront gradient by a measurement matrix, as shown in Eq. \ref{eq:parameter_measurement_matrix_equation}. This relationship is expressed as
\begin{equation}
    I(\vec{x})=\mathbf{M}(\vec{x})\vec{P},
    \label{eq:parameter_measurement_matrix_equation}
\end{equation}
but instead of a Stokes vector, a parameter vector $\vec{P}$ is used. $\vec{P}$ is defined as
\begin{equation}
    \vec{P}=[S_0, \; S_1,\;S_2,\;S_3,\;\frac{\partial W}{\partial x},\; \frac{\partial W}{\partial y}],
    \label{eq:parameter_vector}
\end{equation}
where $S_i$ are the Stokes parameters, $\frac{\partial W}{\partial x}$ is the wavefront gradient in the x dimension, and $\frac{\partial W}{\partial y}$ is the wavefront gradient in the y dimension. The monochromatic measurement matrix is constructed similarly as Eq. \ref{eq:measurement_matrix_construction} and the unknown parameter vector $\vec{P}_U$ is retrieved similarly as Eq. \ref{eq:Stokes_retrieval} and are expressed in Eqs. \ref{eq:parameter_measurement_matrix_construction} and \ref{eq:Parameter_retrieval}.
\begin{equation}
    \mathbf{M}(\vec{x})=[I_{\vec{x},1}, \; I_{\vec{x},2}, \; \cdots , \; I_{\vec{x},N}] 
    \begin{bmatrix}
        S_{0,1}, \; S_{0,2}, \; \cdots , \; S_{0,N} \\
        S_{1,1}, \; S_{1,2}, \; \cdots , \; S_{1,N} \\
        S_{2,1}, \; S_{2,2}, \; \cdots , \; S_{2,N} \\
        S_{3,1}, \; S_{3,2}, \; \cdots , \; S_{3,N} \\
        \frac{\partial W_1}{\partial x},\;\frac{\partial W_2}{\partial x},\; \cdots,\; \frac{\partial W_N}{\partial x} \\
        \frac{\partial W_1}{\partial y},\;\frac{\partial W_2}{\partial y},\; \cdots,\; \frac{\partial W_N}{\partial y} \\
    \end{bmatrix}^{-1}
    \label{eq:parameter_measurement_matrix_construction}
\end{equation}
\begin{equation}
   \vec{P}_{U}=\mathbf{M}(\vec{x})^{-1} I_{meas}(\vec{x})
    \label{eq:Parameter_retrieval}
\end{equation}

\section{Materials and methods}
\subsection{Experimental setup}
\begin{figure}
    \centering
    \includegraphics[width=\linewidth]{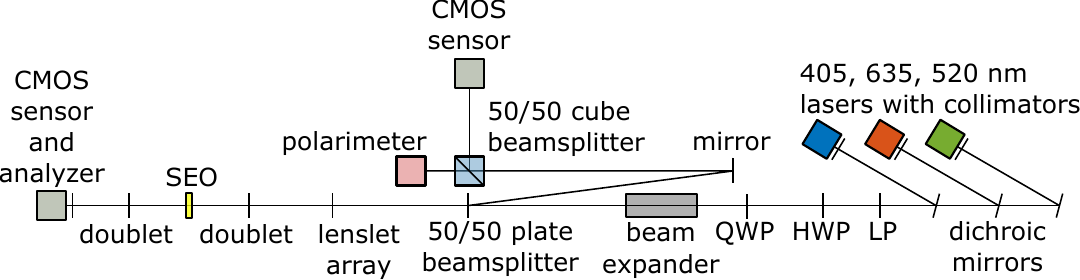}
    \caption{Schematic of the SHWFS-STIP experimental setup, including the preparation and measurement of the beam's polarization state and wavefront gradient.}
    \label{fig:experimental_schematic}
\end{figure}
The setup as shown in Fig. \ref{fig:experimental_schematic} was constructed. Three fiber-coupled benchtop lasers at 405, 520, and 630 nm each with air-spaced doublet fiber collimators for each wavelength are used as individual input sources. The three beams are combined into a common path using dichroic mirrors. The polarization state of the input is prepared with a linear polarizer, an achromatic quarter-wave plate (ThorLabs AQWP10M-580), and an achromatic half-wave plate (ThorLabs AHWP10M-580). The beam passes through a beam expander and then is incident on a 50/50 plate beamsplitter. The reflected beam is sent to instruments for measuring the polarization and the wavefront tilt gradient states. The polarization state is measured with a polarimeter (ThorLabs PAX1000VIS) and the gradient state is measured with a CMOS image detector (FLIR BFLY-U3-05S2C). The transmitted beam continues to the SHWFS-STIP. The lenslet array used is a mounted fused silica lenslet array (ThorLabs MLA150-7AR-M) and is used to create an array of point sources for the star test imaging polarimeter. For the star test imaging polarimeter, a pair of 25.4 mm diameter achromatic doublets with a focal length of 125 mm were used to create the 4F imaging system. A hydraulic pressure SEO\cite{spiecker_stress_2025} and a circular aperture with a radius that is 20\% the SEO's radius (2.54 mm) was placed at the Fourier plane of the 4F system. A CMOS image detector (Basler acA5472-17um) with a right-circular analyzer in front of it was placed at the image plane of the 4F system.

\subsection{Measurement of generated polarization states and wavefront gradients}
\begin{figure}
    \centering
    \includegraphics[width=0.75\linewidth]{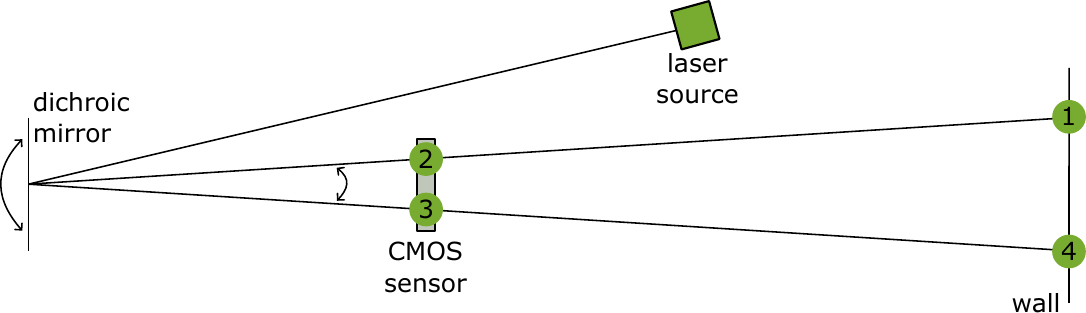}
    \caption{Schematic describing how the wavefront gradient is calibrated on the gradient state sensor. Numbered spots indicate the order where the spot location was recorded. The distance between spots 1 and 4 and the propagation distance are compared to the distance between spots 2 and 3 and the propagation distance to verify that the same angle was measured. The detector is added after spot 1 is recorded and removed after spot 3 is recorded.}
    \label{fig:wavefront_gradient_calibration}
\end{figure}

Equation \ref{eq:parameter_measurement_matrix_construction} shows that in order to be able to construct the monochromatic measurement matrix, the input's wavefront gradient and polarization state and the corresponding output image must be known. The polarization state is simply measured with a polarimeter, which provides the measured Stokes parameter values for S0, S1, S2, and S3. A tilted wavefront gradient was generated by rotating the dichroic mirror that the input beam is incident on. To measure the angle of rotation of the dichroic mirror, which corresponds to a tilt wavefront gradient that is incident on the lenslet array, an aperture (placed immediately after the beam expander, using a flip mount) was used to generate a pencil beam. The pencil beam is incident on the gradient state CMOS image detector, allowing for an analysis of the spot location and displacement relative to a reference location. The wavefront gradient generated is determined with
\begin{equation}
    \theta_{x,y}=dW / dx,y=\sin^{-1} (x,y / z),
    \label{eq:detected_wavefront_gradient}
\end{equation} where x,y is the transverse spot displacement in orthogonal directions as measured on the detector, and z is the propagation distance as measured with a tape measure. To determine the spot location on the sensor, a gaussian filter was used on the recorded image, then the image is thresholded by 50\% of the peak irradiance, and then the location identified by center-of-mass centroiding. The wavefront gradient measured by gradient state sensor is calibrated by measuring the spot displacement produced by a single rotation of the dichroic mirror on a wall as shown in Fig. \ref{fig:wavefront_gradient_calibration}.
An "unit" tilt gradient was chosen to be 200 $\mu$rad and was chosen based on constraints on the active detection area on the gradient state sensor and spot displacement. An efficient sub-pixel image registration algorithm authored by Guizar-Sicairos, Thurman, and Fienup\cite{guizar-sicairos_efficient_2008} (referred to as dftregistration in this paper, available at MathWorks File Exchange\cite{noauthor_efficient_2025}) was used to determine the displacement of the PSF and to quantify changes to the PSF from the reference. The sub-pixel image registration by cross correlation algorithm is based on the idea that a Fourier transform of a translated version of an image contains a global complex phase factor which can be recovered by phase retrieval. The error metric used is a normalized root mean square error quantified through the use of a normalized cross correlation of two images. This dftregistration algorithm was used to determine the amount of displacement each PSF undergoes on the SHWFS-STIP sensor for a given tilt gradient. 

\subsection{Calibration of the SHWFS-STIP}
Since each point source produced by the lenslet array has its own PSF, each lenslet requires its own measurement matrix to be constructed. The measurement matrix is constructed in accordance to Eq. \ref{eq:parameter_measurement_matrix_construction}, using reference states and tested with a combination of reference states and states not utilized in the construction of the measurement matrix. The experimental dataset contains nine polarization states that were generated and for each polarization state, 11 gradient states were generated. Of the nine polarization states generated, six were used in the construction of the measurement matrix and three were not. For each generated polarization state, the same nine gradient states were used in the construction of the measurement matrix and same two were not. In testing the performance of the parameter vector retrieval, a test set of polarization states was chosen from the experimental dataset. Two polarization states that were included in the construction of the measurement matrix and three that were not included were used. For each polarization state, all 11 of the recorded gradient states were used. In this test set, there are a total of 99 unique polarization and gradient state combinations. Since each PSF has its own measurement matrix, the angular error of the Stokes vector retrieval of each PSF is determined and used in statistical analysis. The retrieval of the wavefront gradient is evaluated through wavefront reconstruction with Zernike polynomials and their associated coefficients for each input gradient state. The calculated Zernike coefficients are compared against theoretical values for accuracy in wavefront reconstruction.

\section{Results}
\subsection{Calibration and Measurement }
For each irradiance measurement performed, the recorded image utilizes the same 3,332 PSFs for analysis across 55 test states combining different polarization and gradient states. Table \ref{tab:pixel_shift_gradient} shows the calibrated pixel shift as determined by the dft registration algorithm for four reference wavefront gradient states. Calibration measurements yield a mean pixel shift of $-$.02 for no wavefront gradient and 0.48 for 200 $\mu$rad of wavefront gradient. The root mean square variation for the PSF displacement measurements is 0.4 pixels in all cases, with 79,968 PSFs from reference states used in the construction of the monochromatic measurement matrix. The results of the calibrated pixel shift agree within error the predicted pixel shift of 0.43 pixels for a 200 $\mu$rad of wavefront gradient (as determined by the lenslet's effective focal length (5.2 mm) and the pixel pitch on the sensor (2.4 $\mu$m)). 
\begin{table}
    \centering
    \begin{tabular}{cccc}
        $\frac{dW}{dx}$ & $\frac{dW}{dy}$ & Displacement measurement & Zero measurement\\
        $(\mu\text{rad})$ & $(\mu\text{rad})$ & (pixels) & (pixels)\\
        \hline
       -200 & 0 & -0.48 & -0.04 \\
        200 & 0 & 0.47 & 0.00  \\
        0 & -200 & -0.47  & 0.00 \\
        0 & 200 & 0.51 & -0.02  \\
    \end{tabular}
    \caption{Calibrated pixel shift for four of the reference gradient states. The amount of pixel shift was determined with use of the dftregistration algorithm for N=19,992 PSFs for each reference gradient state. (A reference gradient state uses 3332 PSFs from each of the six reference polarization states.). The root mean square variation for the PSF displacement measurements is 0.4 pixels in all cases.}
    \label{tab:pixel_shift_gradient}
\end{table}

\begin{figure}
    \centering
    \includegraphics[width=1\linewidth]{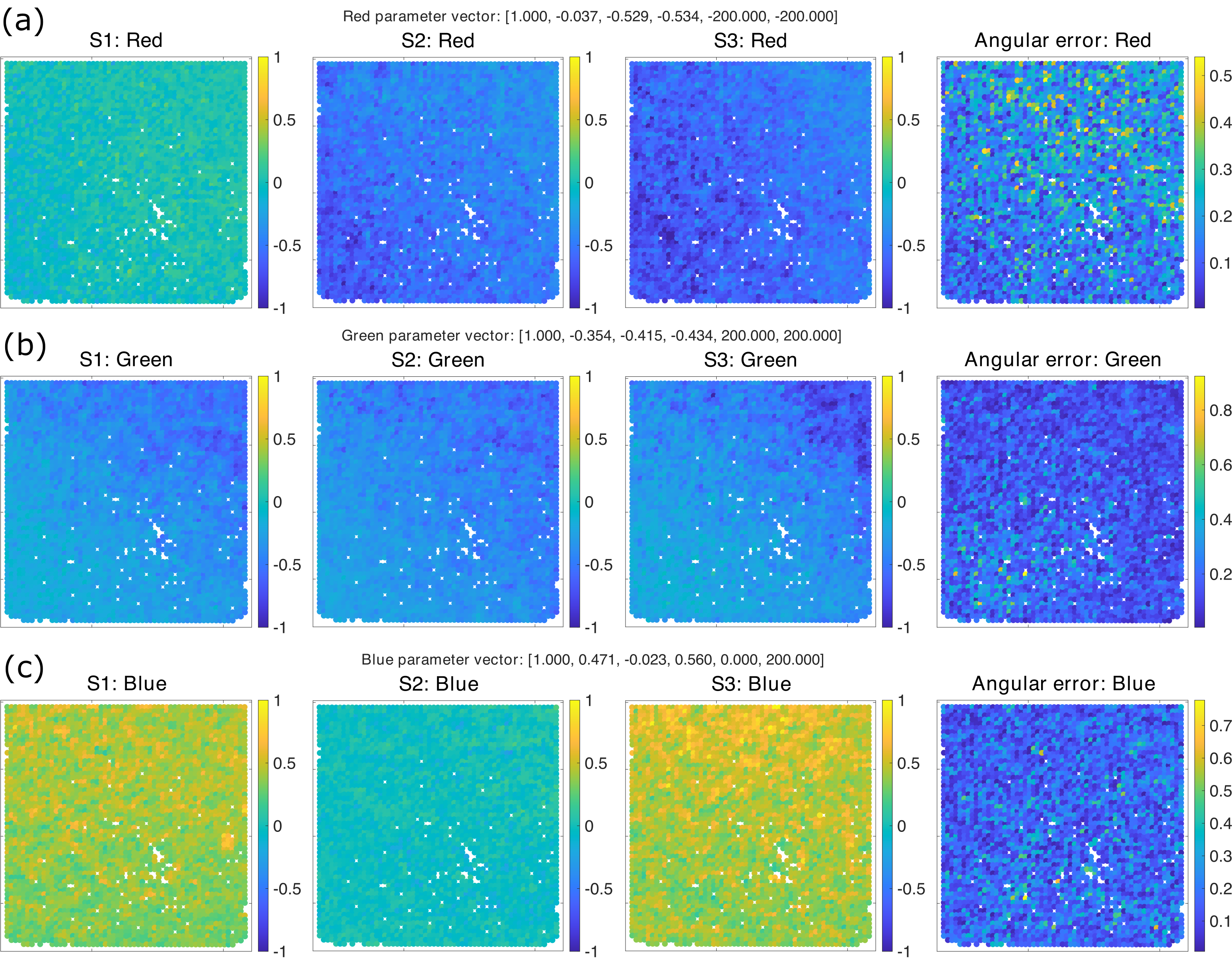}
    \caption{Spatially resolved Stokes parameter retrieval and angular error (measured in rad) at $c_{ref} = 2.88\pi$. (a) Red input with a Stokes vector of [1, -0.037, -0.529, -0.534]. (b) green input with a Stokes vector of [1, -0.345, -0.415, -0.434]. (c) Blue input with a Stokes vector of [1, 0.471, -0.023, 0.560].}
    \label{fig:spatially_resolved_stokes_green}
\end{figure}

\begin{figure}
    \centering
    \includegraphics[width=\linewidth]{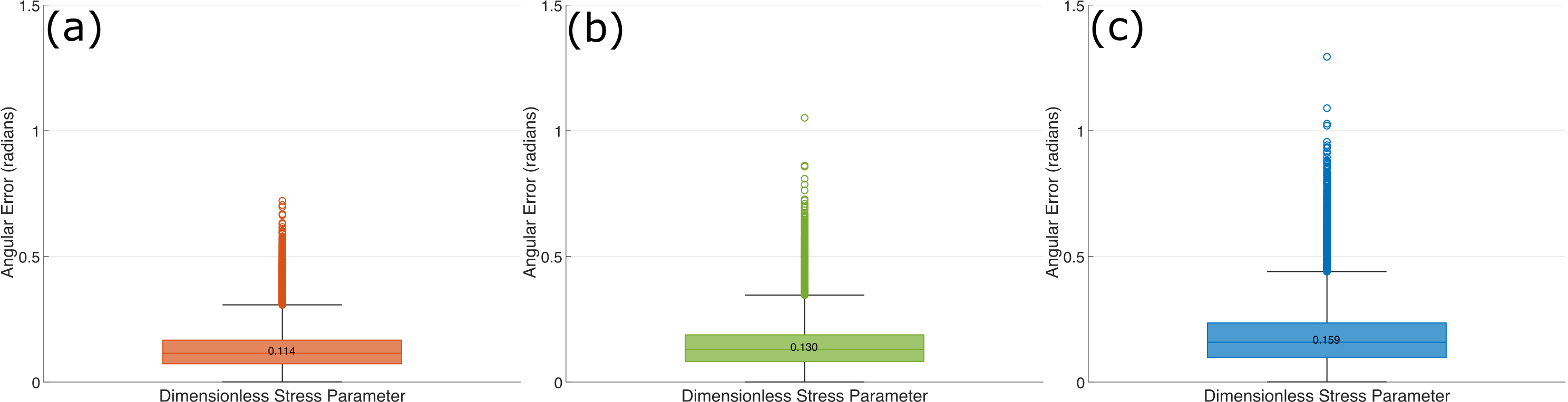}
    \caption{Box plot of angular error in the Stokes vector retrieval. The number in the middle of the box indicates the median value of the angular error. (a) Red input with a median value of 0.114 rad. (b) Green input with a median value of 0.130 rad. (c) Blue input with a median value of 0.159 rad.}
    \label{fig:box_plot_angular_error}
\end{figure}

Figure \ref{fig:spatially_resolved_stokes_green} shows a sample result of spatially resolved Stokes vector retrieval for each of the input wavelengths. 
 The angular error is determined for each individual PSF (N=183,260 PSFs from all states in the test set for a single input wavelength). Figure \ref{fig:box_plot_angular_error} shows the statistics of the angular error in Stokes parameter retrieval for each of the input wavelengths. Results show that red has a median angular error of 0.117 rad, green has a median angular error of 0.133 rad, and blue has a median angular error of 0.167 rad. 

\begin{figure}
    \centering
    \includegraphics[width=1\linewidth]{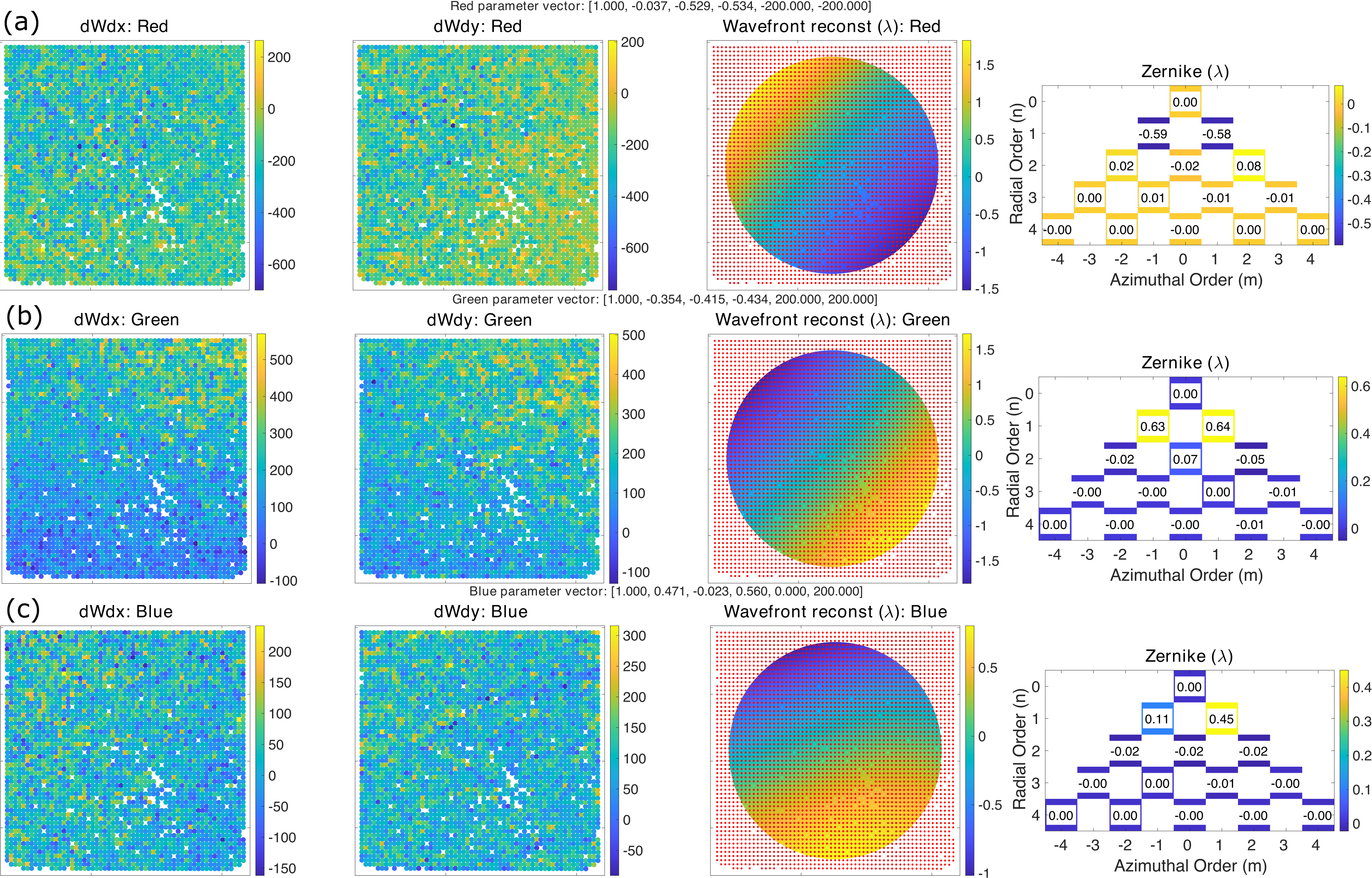}
    \caption{Spatially resolved wavefront gradient retrieval (measured in $\mu\text{rad}$) and wavefront reconstruction using Zernike polynomials. From left to right: the wavefront gradient in the horizontal direction, the wavefront gradient in the vertical direction, the wavefront reconstruction (red dots indicate locations of utilized PSFs for retrieval, only the points within the radius of the reconstructed wavefront are used in the reconstruction), and the Zernike coefficients (measured in waves) for the corresponding radial and azimuthal orders are shown. (a) Red input with $dW/dx = -200 \;\mu$rad and $dW/dy = -200 \;\mu$rad. (b) green input with $dW/dx = 200 \;\mu$rad and $dW/dy = 200 \;\mu$rad. (c) Blue input with $dW/dx = 0 \;\mu$rad and $dW/dy = 200 \;\mu$rad.}
    \label{fig:spatially_resolved_wavefront_green}
\end{figure}

\begin{figure}
    \centering
    \includegraphics[width=\linewidth]{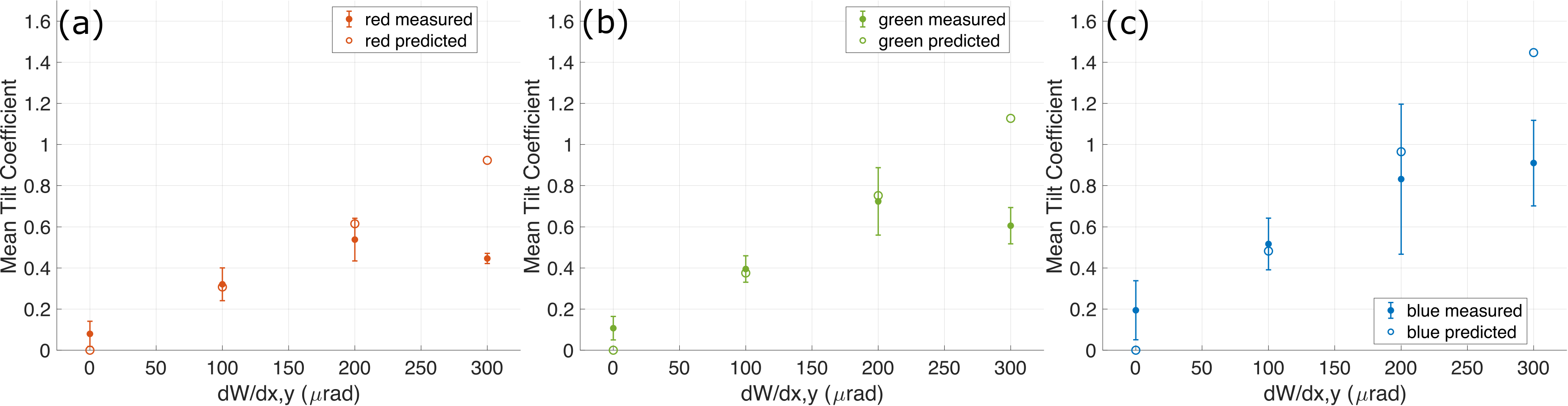}
    \caption{Mean value of the fitted Z02 and Z03 Zernike coefficients (in waves) and the standard deviation for wavefront gradients of 0, 100, 200, and 300 $\mu\text{rad}$. (a) Red input. (b) Green input. (c) Blue input.}
    \label{fig:Zernike_coeff_stats}
\end{figure}

Figure \ref{fig:spatially_resolved_wavefront_green} shows a sample result of spatially resolved wavefront gradient retrieval and wavefront reconstruction for each of the input wavelengths. 
The Z02 and Z03 Zernike coefficients retrieval uses the complete set of PSFs utilized for each test state (N=55 test cases for a single input wavelength). Figure \ref{fig:Zernike_coeff_stats} shows the statistics of the combined Z02 and Z03 Zernike coefficients used to evaluate the wavefront gradient retrieval for each of the input wavelengths. 

\subsection{SHWFS-STIP measurement}
\begin{figure}
    \centering
    \includegraphics[width=1\linewidth]{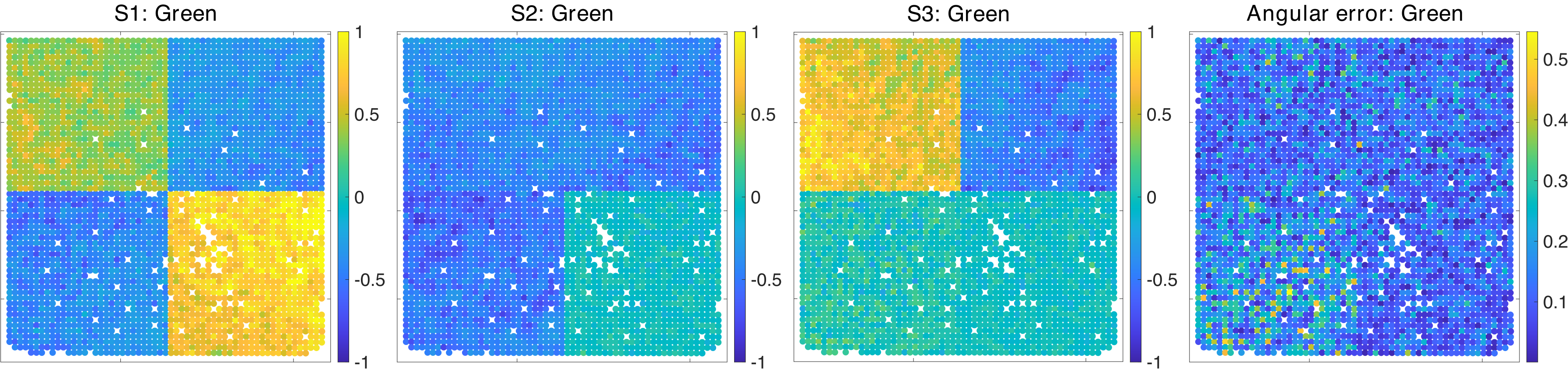}
    \caption{Spatially resolved Stokes parameter retrieval of a virtual measurement and angular error (measured in rad).}
    \label{fig:composite_stokes_ret}
\end{figure}

\begin{figure}
    \centering
    \includegraphics[width=0.35\linewidth]{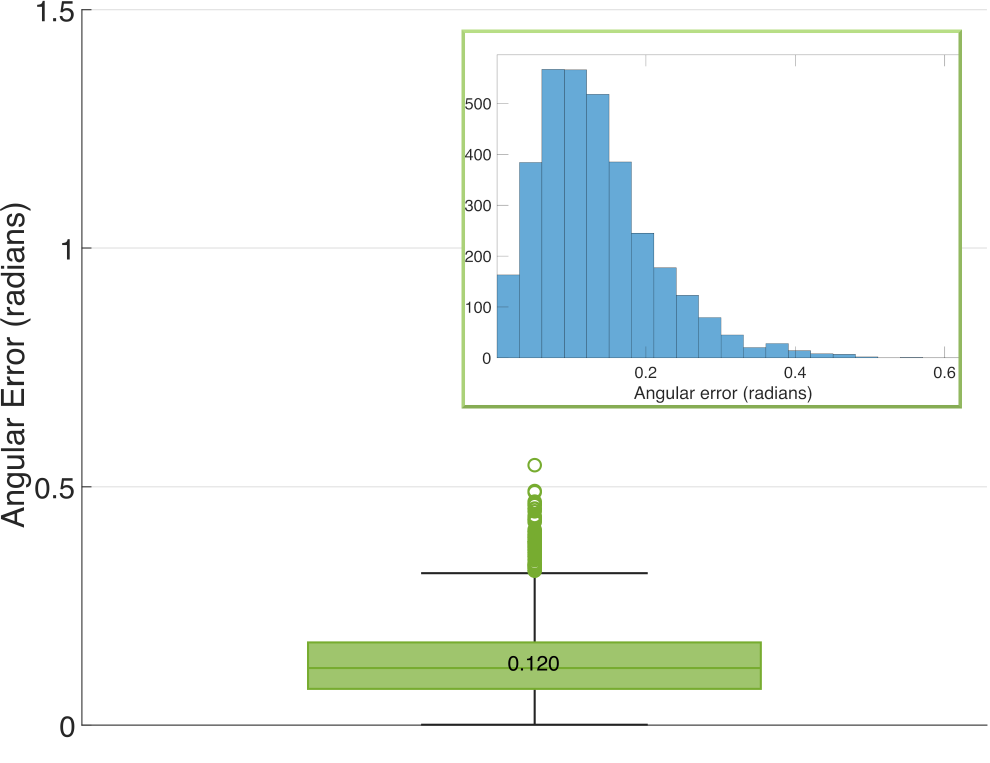}
    \caption{Box plot of angular error in Stokes vector retrieval for the virtual measurement for a green input. The number in the middle of the box indicates the median value of the angular error which is 0.120 rad. An inset displays the same plotted data as a histogram.}
    \label{fig:composite_box_plot}
\end{figure}

\begin{figure}
    \centering
    \includegraphics[width=0.75\linewidth]{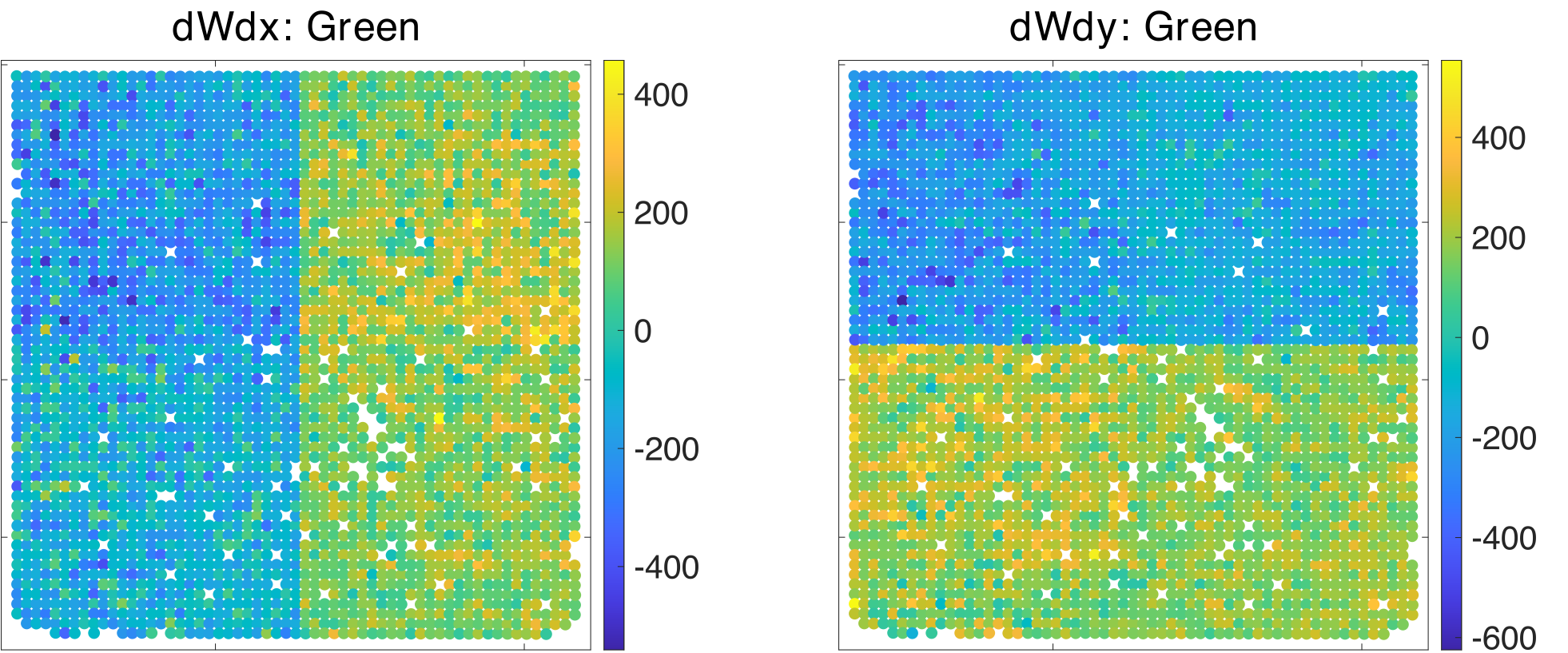}
    \caption{Spatially resolved wavefront gradient retrieval (measured in $\mu$rad) of the virtual measurement.}
    \label{fig:composite_gradient_ret}
\end{figure}

\begin{figure}
    \centering
    \includegraphics[width=1\linewidth]{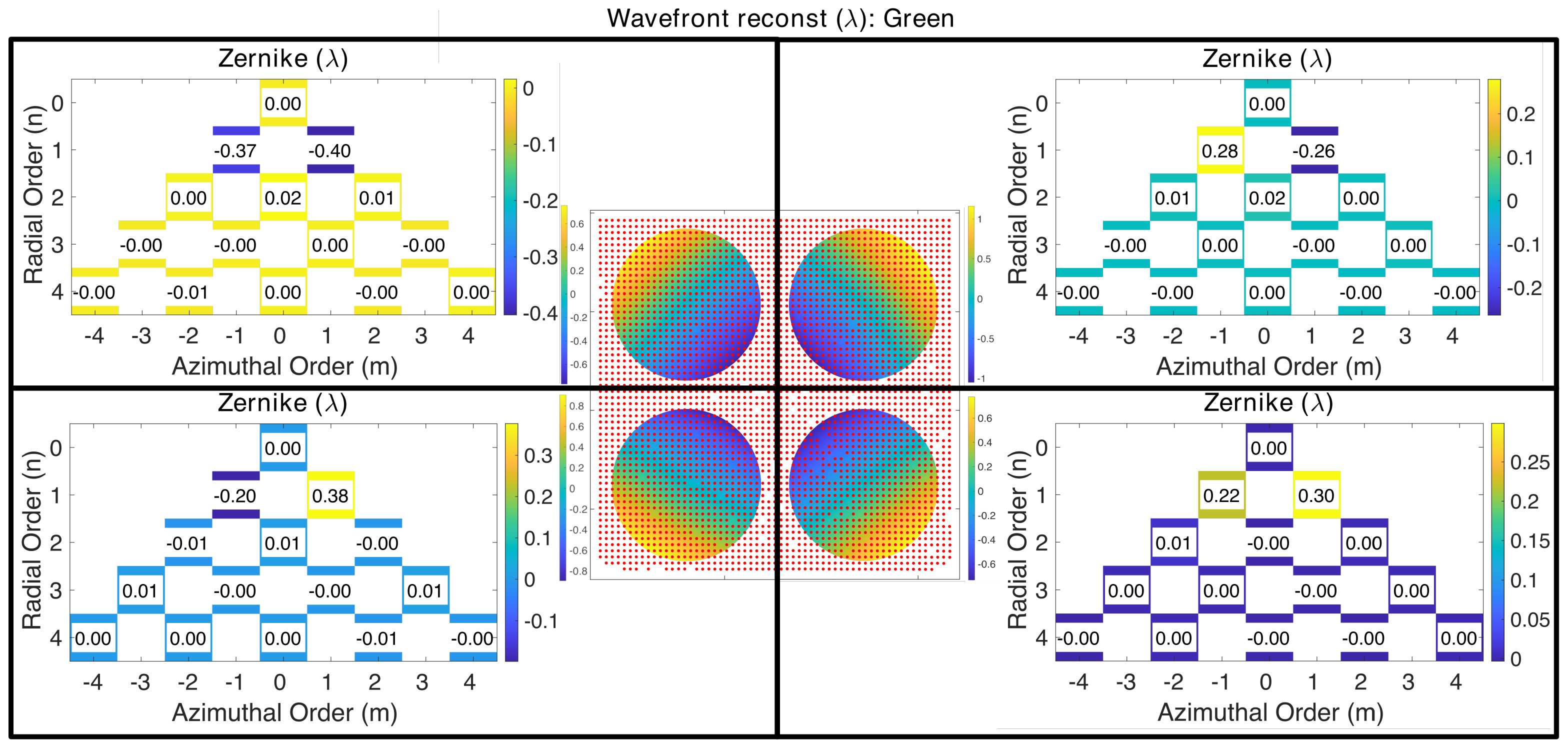}
    \caption{Wavefront reconstruction (broken down into quadrants, measured in waves) of the retrieved wavefront gradient shown in Fig. \ref{fig:composite_gradient_ret}. The predicted Z02 and Z03 coefficients is 0.36 $\lambda$ (with appropriate signs belonging to each quadrant). Retrieved values agree with the predicted Z02 and Z03 coefficient values within error (from calibration for 200$\mu$rad: $\pm$0.16 $\lambda$) for all quadrants.}
    \label{fig:composite_wavefront_reconstruction}
\end{figure}

To test the SHWFS-STIP and parameter vector retrieval, a virtual measurement was created by using a composite image of four recorded states. A different quadrant was taken from each of the four recorded state to produce a new composite image to be used in the parameter vector retrieval. The quadrants are numbered from left to right, top to bottom. The first quadrant has a parameter vector of [1, 0.3478, -0.3636, 0.6304, -200, -200], the second quadrant has a parameter vector of [1, -0.3376, -0.4126, -0.4272, 200, -200], the third quadrant has a parameter vector of [1, -0.4409, -0.5108, 0.0045, -200, 200], and the fourth quadrant has a parameter vector of [1, 0.9095, -0.0267, 0.0000, 200, 200]. Figure \ref{fig:composite_stokes_ret} shows the Stokes parameter retrieval and the angular error of this virtual measurement, and Fig. \ref{fig:composite_box_plot} shows a box plot of the angular error with a median value of 0.120 rad.
Figure \ref{fig:composite_gradient_ret} shows the retrieved wavefront gradient of the virtual measurement and Fig. \ref{fig:composite_wavefront_reconstruction} shows the reconstructed wavefront (for each quadrant), demonstrating successful wavefront gradient retrieval within error for all quadrants.
 
\section{Discussion}
\subsection{Polarimetry}
Results presented in this paper shows that polarimetry in SHWFS-STIP performs well with angular errors in the order of 100 mrad and is consistent across input wavelengths. The virtual measurement shows angular error values similar to the angular error seen in calibration, which is expected since polarimetry is done on a PSF-by-PSF basis. Those results are compared to previous results\cite{spiecker_single-shot_2025} for monochromatic Stokes vector retrieval and polychromatic Stokes vector retrieval for $c_{ref}=2.88\pi$. Previous results showed that the median monochromatic angular error for red, green, and blue inputs respectively are 0.038, 0.048, and 0.053 rad, while the median polychromatic angular error for red, green, and blue inputs respectively are 0.082, 0.153, and 0.069 rad. Comparing previous Stokes vector retrieval performance with the results obtained in this paper show a small increase in angular error whenever more information is encoded in the irradiance patterns. In evaluating the performance of the Stokes vector retrieval through the use of the angular error, results suggest that additional spectral information creates more angular error than the presence of a wavefront gradient. 

\subsection{Wavefront sensing}
The work discussed in this paper demonstrates a good test case for the sensitivity of SHWFS-STIP in detecting small wavefront gradients. It has been shown that the system is capable of detecting gradients as small as 100 $\mu$rad, corresponding to a PSF shift of approximately a quarter of a pixel in this system configuration. However, this is in part limited by our ability to accurately induce (and independently measure) a smaller linear gradient.  
To assess the potential performance of the system for smaller wavefront gradients we can examine the recovered Z02 and Z03 Zernike coefficients against the expected values, and compare them with the residuals from other low-order Zernikes. The error from the expected values is of order $\pm0.1\lambda$, but it is notable that the residuals for other Zernike coefficients are much smaller than this value, indicating that the error is linked not only to the noise-limited retrieval accuracy (with residuals of $0.01\lambda$ and smaller) but also to the accuracy of the calibration.      

\begin{figure}
    \centering
    \includegraphics[width=0.75\linewidth]{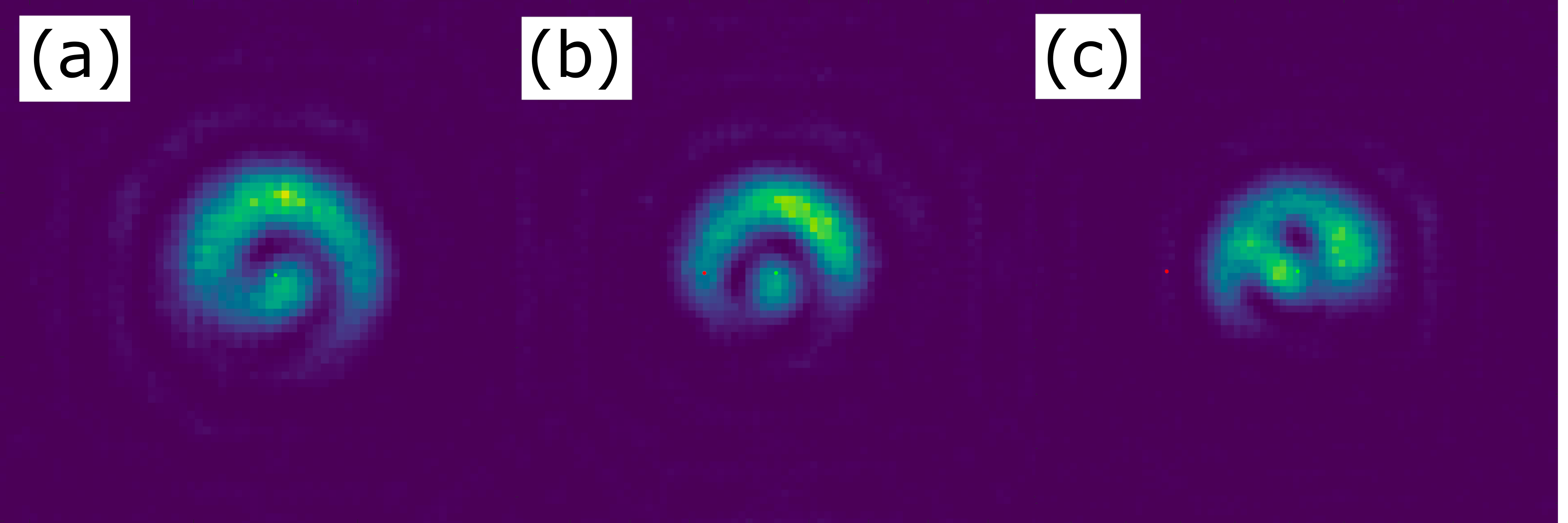}
    \caption{Wavefront gradient dependence of the PSF of a V polarized input. The red dot indicates the reference position (0 $\mu$rad gradient) and the green dot indicates the center of the PSF. (a) 0 $\mu$rad gradient, (b) 400 $\mu$rad gradient in the positive horizontal direction, and (c) 800 $\mu$rad gradient in the positive horizontal direction.}
    \label{fig:gradient_dependency}
\end{figure}

For higher gradients we find that, for each input wavelength, the recovered Z02 and Z03 coefficients increase linearly with increasing values of the input wavefront gradient as expected. The recovered coefficients fall within experimental error with the notable exception of wavefront gradients of 300 $\mu$rad where the retrieved Z02 and Z03 Zernike coefficients are underestimated compared to the directly measured gradients. It is known through this work (and previous work) that the irradiance pattern has a polarization state dependency. However, simulations with increasing wavefront gradients suggest that the irradiance pattern also has a wavefront gradient state dependency as shown in Fig. \ref{fig:gradient_dependency}. 
The likely reason for this is that, in our system, increasing gradients are accompanied by a shift in the irradiance pattern at the SEO plane. This displacement breaks the rotational symmetry of the retardance and fast axis orientations present in the SEO that the irradiance pattern is incident upon, causing irregular but predictable changes to the resulting PSF. It is anticipated that correcting this would require additional wavefront gradient states to be included in the set of reference states in constructing the measurement matrix.

\subsection{Dynamic range and system limitations of SHWFS-STIP}
The work presented in this paper demonstrates the performance of SHWFS-STIP with wavefront gradients as low as 100 $\mu$rad, with a maximum wavefront gradient of 300 $\mu$rad. Constraints on the current experimental setup prevented larger wavefront gradients from being generated but this is not an inherent limitation of SHWFS-STIP. Developing the ability to generate larger wavefront gradients will enable the exploration of the crosstalk between adjacent PSFs due to their respective displacements and how the performance of wavefront sensing is impacted. With typical Shack-Hartmann sensing, this is a centroid localization problem. With the unique irradiance patterns that are dependent on input parameters present in SHWFS-STIP, it is expected that parameter retrieval using the techniques discussed in this paper will be affected by overlap between adjacent PSFs due to both the confusion of separating irradiance patterns and in accounting for the mutual (spatial) coherence between adjacent lenslets. 

Simulations also suggest that the PSF, when imaged through an SEO can have a dependence on the wavefront gradient that is independent of the polarization state dependence of the PSF; future work should determine the required number of of reference states necessary to construct a measurement matrix that can accurately retrieve the wavefront gradient within a given range of gradients. Additionally, future work should develop inputs with complex wavefront gradients in addition to complex polarization patterns to test the upper limit of SHWFS-STIP and to determine if there are any unexpected interactions between polarization and gradient states. Zimmerman and Brown\cite{zimmerman_star_2016} demonstrated that it is possible to perform real-time polarimetry with the star test imaging polarimeter. Since the basic techniques used are the same, it is theoretically possible to enable real-time polarimetry and wavefront sensing with low computational requirements.
The dynamic range of a SHWFS-STIP measurement carries the usual limitations of a Shack Hartmann sensor -- at its most fundamental it can be defined as the ratio of the maximum 'deflection' of a PSF due to the measured wavefront gradient to the minimum PSF displacement that is measurable by a given system.  For an ideal (e.g. Gaussian) or space-invariant focal spot, this is a localization problem that can be solved using Cramer-Rao bounds \cite{thomas_optimized_2004, wei_analysis_2020}. For our system, the problem is somewhat more complicated since the PSF itself is polarization sensitive and, as noted, can have a shape that varies with a wavefront gradient. Nevertheless, it is possible to assess an empirical dynamic range using the earlier cited localization uncertainty as a lower limit and the lenslet pitch as an upper limit to estimate a likely dynamic range greater than 250 for the current system.   

\section{Future Directions}

With the increasing interest in complex light fields with exotic polarization states (sometimes known as structured beams), interest in performing simultaneous spatially resolved polarimetry and wavefront sensing has grown. Utilizing similar properties of a lenslet array and stress birefringence, Wakayama, et. al. \cite{wakayama_simultaneous_2024} demonstrated the ability to characterize radially and azimuthally polarized fields with and without optical vortices. Fields where orthogonal components contain several frequencies and are phase locked, the electric field vector traces out a Lissajous figure and are called a Lissajous state. Polarization singularities in a polychromatic (or in the simplest case, a bichromatic) vector field create a Lissajous singularity. \cite{kessler_lissajous_2003, miao_design_2022} To characterize such fields, it is necessary to be able to perform simultaneous spatially resolved spectropolarimetry and wavefront sensing. Previous work by Spiecker and Brown\cite{spiecker_single-shot_2025} demonstrated that it is possible to retrieve the spatially resolved Stokes vector of each input wavelength contained in an input made of a superposition of multiple wavelengths. This paper has demonstrated that it is possible to retrieve a wavefront gradient based on the PSF displacement contained in the recorded image. The use of the parameter vector in the construction of the monochromatic measurement matrix is compatible with the construction of the polychromatic measurement matrix as discussed by Spiecker and Brown\cite{spiecker_single-shot_2025}.  Combining the use of the parameter vector and the construction of a polychromatic measurement matrix makes it possible to enable single-shot, simultaneous spatially resolved spectropolarimetry and wavefront sensing using SHWFS-STIP.

\subsection*{Disclosures}
The authors declare that there are no financial interests, commercial affiliations, or other potential conflicts of interest that could have influenced the objectivity of this research or the writing of this paper.

\subsection* {Code, Data, and Materials Availability} 
Data underlying the results presented in this paper are not publicly available at this time but may be obtained from the authors upon reasonable request.

\subsection* {Acknowledgments}
We want to acknowledge previous work in modeling and polarimetry with SEOs by the following former members of, and collaborators with, the Brown lab: Alexis Vogt, Amber Beckley, Roshita Ramkhalawon, Ashan Ariyawansa, Brandon Zimmermann, and Miguel Alonso. Likewise, we want to acknowledge people who have provided valuable advice and discussion: Brian Kruschwitz and Katelynn Bauer. Furthermore, we acknowledge financial support from: the Horton Fellowship from the Laboratory for Laser Energetics, the GAANN fellowship from the US Department of Education (P200A210035-322 23), the Institute of Optics at University of Rochester, and SPIE.


\bibliography{references}   
\bibliographystyle{spiejour}   

\listoffigures
\listoftables

\end{spacing}
\end{document}